\documentclass[aps,prd,twocolumn,10pt]{revtex4-1} 
\pdfoutput=1
\usepackage{graphicx}
\usepackage{amssymb, amsmath, amsthm, amsfonts, bm, mathrsfs, wasysym, txfonts, bbm, enumerate, color}

\begin{document}
\title[Relativistic superfluidity]{Relativistic superfluidity and vorticity from the nonlinear Klein-Gordon equation}

\author{Chi Xiong$^{1}$, Michael R.R. Good$^{1}$, Yulong Guo$^{2}$, Xiaopei Liu$^{3}$, and Kerson Huang$^{1, 4}$}
\email{Email: xiongchi@ntu.edu.sg \\Email: mgood@ntu.edu.sg \\Email: gyl.this@gmail.com \\Email: liuxp@ntu.edu.sg \\Email: kerson@mit.edu}
\affiliation{$^{1}$Institute of Advanced Studies, Nanyang Technological University, Singapore 639673}
\affiliation{$^{2}$School of Computer Science and Engineering, South China University of Technology, Guangzhou, China 510641}
\affiliation{$^{3}$School of Computer Engineering, Nanyang Technological University, Singapore 639673}
\affiliation{$^{4}$Physics Department, Massachusetts Institute of Technology, Cambridge, MA, USA 02139}


\begin{abstract}
We investigate superfluidity, and the mechanism for creation of quantized
vortices, in the relativistic regime. The general framework is a nonlinear
Klein-Gordon equation in curved spacetime for a complex scalar field, whose
phase dynamics gives rise to superfluidity. The mechanisms discussed are local
inertial forces (Coriolis and centrifugal), and current-current interaction
with an external source. The primary application is to cosmology, but we also
discuss the reduction to the non-relativistic nonlinear Schr\"{o}dinger
equation, which is widely used in describing superfluidity and vorticity in
liquid helium and cold-trapped atomic gases.

\vspace{0.5cm}



\end{abstract}
\maketitle

\section{Introduction}

In quantum mechanics, a system of particles is describable by a complex wave
function, which has a modulus as well as a phase, and the existence of the
quantum phase is an essential distinction between quantum mechanics and
classical mechanics. Macroscopic phase coherence (correlation of the quantum
phase over macroscopic distances) gives rise to superfluidity and
occurs in Bose-Einstein condensates (BEC) in diverse systems \cite{Khalatnikov, RMP, RMP2}: 
liquid $^{4}$He at low temperatures; cold trapped atomic gases; central region in relativistic heavy-ion collisions; 
Higgs or Higgs-like fields over cosmological scales; 
superconducting metals; liquid $^{3}$He at low temperatures; interior of neutron stars.
The last three refer to superconductivity, which can be viewed as
the superfluidity of condensed fermion pairs. Ginzburg and Landau \cite{GL} propose a
general phenomenological theory that describes the phase coherence in terms of
a complex scalar field, which is viewed as an order parameter that emerges in
a phase transition below a critical temperature. This phase transition is
associated with spontaneous breaking of global gauge symmetry, i.e., the
invariance of the wave function under a constant change of phase. We adopt
such an approach here, and, in view of cosmological applications, begin with a
complex scalar field in curved spacetime, and then consider the flat-spacetime
and non-relativistic limits. Our treatment is confined to the neighborhood of
the absolute zero of temperature.

We denote the complex scalar field by
\begin{equation}
\Phi=Fe^{i\sigma},
\end{equation}
which satisfies a nonlinear Klein-Gordon equation (NLKG) of the form
$\Box\Phi+f\left(  \Phi\right)  =0,$ where the d'Alembertian operator
$\Box$ is the generalization of $\partial^{\mu}\partial_{\mu}=\nabla^{2}$
$-\frac{1}{c^{2}}\frac{\partial^{2}}{\partial t^{2}}$ to curvilinear
coordinate frames, and $f$ is a nonlinear function, which contains a potential
that has a minimum at $\Phi\neq0$. Thus, there exists a nonzero vacuum field,
which breaks global gauge symmetry spontaneously. The superfluid velocity is
proportional to the spatial gradient of the phase $\nabla\sigma$, which has
quantized circulation:
\begin{equation}
{\displaystyle\oint\limits_{C}}
\nabla\sigma\cdot d\mathbf{s=}2\mathbf{\pi}n\text{ \ }\left(  n=0,\pm
1,\pm2,\ldots\right)  ,
\end{equation}
The integral extends over a spatial closed loop $C$, and the quantization is a
consequence of the fact that $\Phi$, and hence the phase factor $e^{i\sigma}$,
must be a continuous function in space. The contour $C$ encircles a vortex
line, which may meander in space but must end on itself, forming a closed
loop, or terminate on boundaries. The modulus $F$ must vanish on the vortex
line, where it approaches zero continuously, over a characteristic distance,
the healing length. Thus the vortex line is in actuality a vortex filament with a finite effective radius.

The superfluid velocity $\mathbf{v}_{s}$ is a hydrodynamic quantity that obeys
certain conservation laws. As discussed below, it has the form
\begin{equation}
\frac{\mathbf{v}_{\text{s}}}{c}=\xi_{\text{s}}\nabla\sigma,
\end{equation}
where $\xi_{\text{s}}$ is a correlation length that is generally spacetime
dependent, with $\xi_{\text{s}}\rightarrow\hbar/mc$ in the non-relativistic
limit, where $m$ is a mass scale. The fact that the modulus $F$ must vanish at
the vortex center effectively makes the space non-simply connected, i.e.,
there are closed circuits that cannot be shrunken to zero continuously. This
is why we can have $\nabla\times\mathbf{v}_{\text{s}}\neq0$, even though ${\bf v}_{\text{s}}$ is
proportional to a gradient (at least non-relativistically).

In the large-scale motion of a quantum fluid, as in a classical medium such as
the atmosphere or the ocean, vorticity is ubiquitous, being induced through
different means in different systems. In liquid helium in the laboratory, they
can be created through rotation of the container, or through local heat
perturbations. The latter can create quantum turbulence in the form of a
vortex tangle. In the cosmos, quantized vortices in the background superfluid
can be created by a  rotating black hole, a rotating galaxy, or colliding
galaxies. The big bang era could witness the creation of quantum turbulence.
Our main theme is that all these effects can be traced to a universal
mechanism characterized by operator terms in the NLKG that can be associated
with the Coriolis and the centrifugal force:
\begin{align}
R_{\text{Coriolis}} &  =\frac{2\Omega}{c^{2}}\frac{\partial^{2}}{\partial
\phi\partial t},\nonumber\\
~~~R_{\text{centrifugal}} &  =-\frac{\Omega^{2}}{c^{2}}\frac{\partial^{2}
}{\partial\phi^{2}},
\end{align}
where $\Omega$ is an angular velocity, and $\phi$ the angle of rotation. The
angular velocity may be an externally given constant, or it may be a spacetime
function arising from dynamics, interaction with external sources, or
spacetime geometry.

The various threads of our investigation may be summarized by the chart in
Fig.1. We begin in Sec.2 with a general formulation of the NLKG in general
relativity, and define the superfluid velocity field $\mathbf{v}_{\text{s}}$
in terms of the phase of the field. This is followed in Sec.3 by a reduction
to flat Minkowski spacetime. In Sec.4, we then consider the NLKG in a global
rotating frame for orientation, and exhibit the Coriolis force and centrifugal
force (the top link in Fig.1). This enables us to extract local Coriolis and
centrifugal terms from the curved-space equation, by expansion in the local
angular velocity, (the left to right link in Fig.1). The local rotating frames
associated with the Coriolis force leads to frame-dragging. We then make a
non-relativistic reduction in Sec.5, recovering the nonlinear Schr\"{o}dinger
equation, particularly the form used to describe vortex generation in rotating
BEC (the link to the bottom square in Fig.1). In Sec.6, we discuss the
introduction of an external source via current-current coupling, and note how
it can have the same effect as the Coriolis force (the right link in Fig.1).
Finally, in Sec.7, we show some results from numerical computations.


\begin{figure}
[!htb]
\begin{center}
\includegraphics[
width=0.48\textwidth, natwidth=2800,natheight=1400
]
{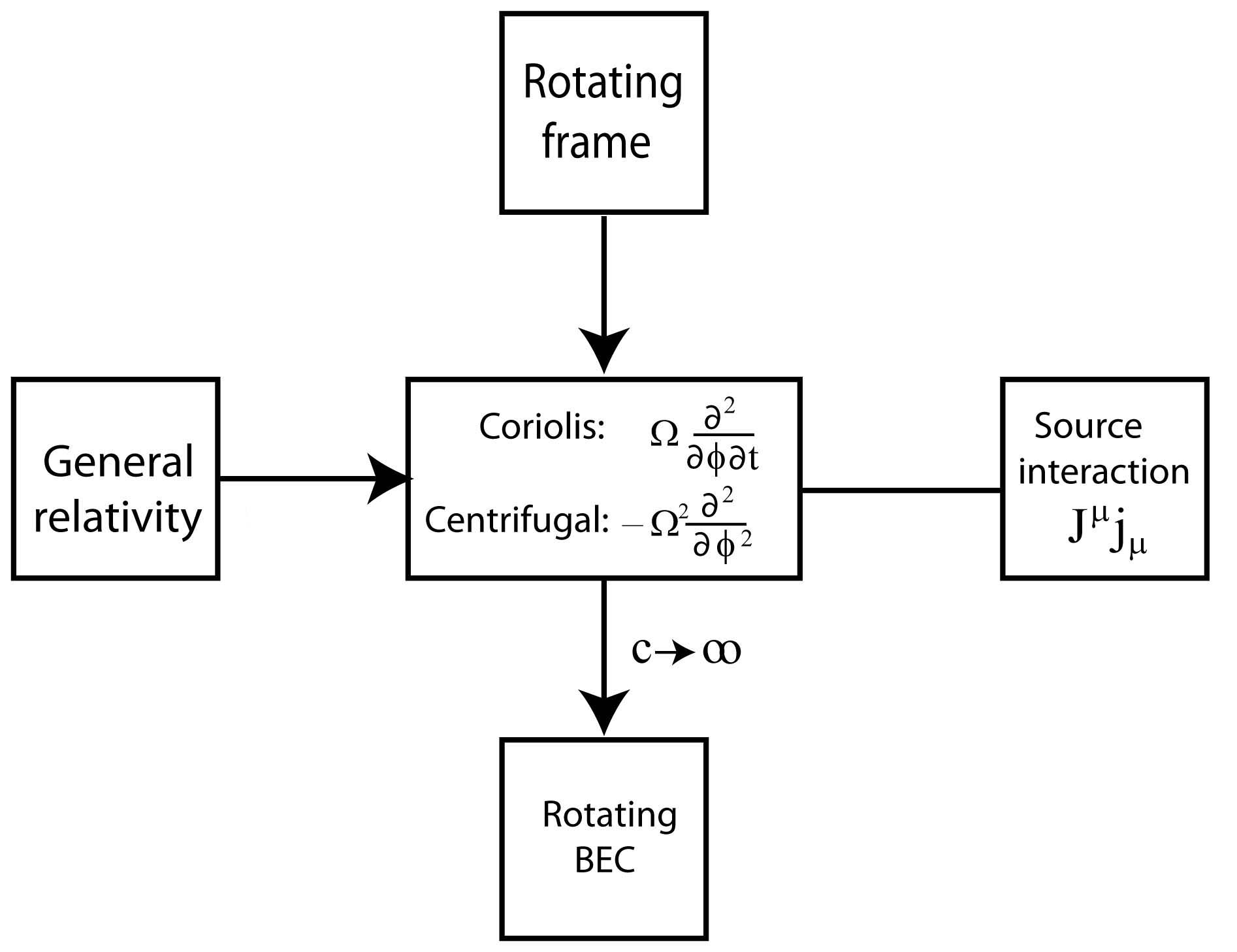}
\caption{Chart showing various threads of our investigation. The general theme,
shown by the central and the left square, is that the mechanisms for vortex
creation are the inertia forces (Coriolis and centrifugal) in rotating
frames, generated by external means or by the spacetime metric, or by external
sources through a current-current interaction. See Sec. I to follow the various threads.}
\end{center}
\end{figure}

\section{Superfluidity in curved spacetime}

We start with the NLKG in a general background metric $g_{\mu\nu}~(\mu
,\nu=0,1,2,3)$. The action of the complex scalar field is given by
\begin{equation}
S=-\int dt\,d^{3}x\sqrt{-g}\left(  g^{\mu\nu}\partial_{\mu}\Phi^{\ast}
\partial_{\nu}\Phi+V\right), \label{phiaction}
\end{equation}
where $g=$ det$\left(  g_{\mu\nu}\right)  $, and $V$ denotes a
self-interaction potential that depends only on $\Phi^{\ast}\Phi$. The
equation of motion is
\begin{equation}
\left(  \Box-V^{\prime}\right)  \Phi=0,\label{sfEOM}
\end{equation}
where $V^{\prime}=dV/d\left(  \Phi^{\ast}\Phi\right)  $, and
\begin{equation}
\Box\Phi\equiv\nabla^{\mu}\nabla_{\mu}\Phi=\frac{1}{\sqrt{-g}}\partial
_{\mu}(\sqrt{-g}g^{\mu\nu}\partial_{\nu}\Phi).
\end{equation}
where $\nabla_{\mu}$ is the covariant derivative. 

In the phase representation $\Phi=Fe^{i\sigma}$, the real and imaginary parts
of (\ref{sfEOM}) lead to the equations:
\begin{align}
\left(  \Box-V^{\prime}\right)  F-F\nabla^{\mu}\sigma\nabla_{\mu}\sigma &
=0,\nonumber\\
2\nabla^{\mu}F\nabla_{\mu}\sigma+F\nabla^{\mu}\nabla_{\mu}\sigma &  =0.
\end{align}
The first equation can be rewritten in the form of a relativistic Euler
equation
\begin{equation}
\nabla_{\mu}\left(  F^{-1}\Box F-V^{\prime}\right)  -2F^{-2}\nabla
^{\lambda}(F^{2}\nabla_{\lambda}\sigma\,\nabla_{\mu}\sigma)=0. \label{KGEuler}
\end{equation}
The second equation is a continuity equation
\begin{equation}
\nabla^{\mu}j_{\mu}=0, \label{consvJ}
\end{equation}
with
\begin{equation}
j_{\mu}\equiv F^{2}\nabla_{\mu}\sigma=F^{2}\partial_{\mu}\sigma.
\label{current}
\end{equation}
There exist a conserved charge
\begin{equation}
Q=\int d^{3}x\sqrt{-g}~j^{0},
\end{equation}
and a covariantly conserved energy-momentum tensor 
\begin{equation}
T_{\mu\nu}=\,\nabla_{\mu}\Phi^{\ast}\nabla_{\nu}\Phi+\nabla_{\nu}\Phi^{\ast
}\nabla_{\mu}\Phi-g_{\mu\nu}\nabla^{\tau}\Phi^{\ast}\nabla_{\tau}\Phi
-g_{\mu\nu}V,
\end{equation}

We \ turn to the definition of the superfluid velocity $\mathbf{v}_{\text{s}}
$. The field $\Phi\left(  \mathbf{x},t\right)  $ corresponds to an order
parameter at absolute zero, and thus represents a pure superfluid. Let $\tau$
be the proper time along a timelike worldline of a fluid element whose
coordinates are
\begin{equation}
x^{\mu}=\left(  c\,t(\tau),~x^{i}(\tau)\right)  ,
\end{equation}
where$~i=1,2,3.$ The superfluid velocity is the 3-velocity of the fluid
element:
\begin{equation}
\mathbf{v}_{\text{s}}=\frac{d\mathbf{x}}{dt}.
\end{equation}
We define a 4-velocity
\begin{equation}
U^{\mu}\equiv\frac{dx^{\mu}}{d\tau}=\left( \gamma c, ~\gamma \mathbf{v}
_{\text{s}}\right)  ,
\end{equation}
where $\gamma\equiv dt/d\tau$. From $ds^{2}=-c^{2}d\tau^{2}=g_{\mu\nu}dx^{\mu
}dx^{\nu},$ we obtain
\begin{equation}
\gamma=\left(  -g_{00}-2g_{0i}\frac{v^i}{c}-g_{ij}\frac{v^{i}v^{j}}{c^{2}
}\right)  ^{-1/2}.
\end{equation}
The superfluid density $\rho_{\text{s}}$ is defined through
\begin{equation}
j^{\mu}=\rho_{\text{s}}U^{\mu}.
\end{equation}
Comparison with (\ref{current}) leads to
\begin{align}
\rho_{\text{s}}  &  = \hbar \left(  c\gamma\xi_{\text{s}}\right)  ^{-1}
F^{2},\nonumber\\
\frac{\mathbf{v}_{\text{s}}}{c}  &  =\xi_{\text{s}}\nabla\sigma
\end{align}
where
\begin{equation}
\xi_{\text{s}}=\left(  \partial^{0}\sigma\right)  ^{-1}.
\end{equation}

\section{Reduction to Minkowski spacetime}

In Minkowski metric $(-1,1,1,1),$ the Euler equation and the continuity
equation become
\begin{align} 
&  \partial_{\mu}\left(  F^{-1}\Box F-V^{\prime}\right)  -2F^{-2}
\partial^{\lambda}(F^{2}\partial_{\lambda}\sigma\,\partial_{\mu}
\sigma)=0,\nonumber\\
&  \partial^{\mu}(F^{2}\partial_{\mu}\sigma)=0.\label{hydro}
\end{align}
The 4-velocity is given by
\begin{equation}
U^{\mu} =\left(  \gamma c,\,\gamma\mathbf{v}_{\text{s}}\right) ,~~~
\gamma =\frac{1}{\sqrt{1-v_{\text{s}}^{2}/c^{2}}}.
\end{equation}
One can check $U^{\mu}U_{\mu}=-c^{2},$ which guarantees $v_{\text{s}}<c$. The
superfluid density and velocity are given by
\begin{align} 
\rho_{\text{s}} &  =-\,\frac{\hbar\dot{\sigma}}{c^{2}\gamma}\,F^{2}
,\nonumber\\
\frac{\mathbf{v}_{\text{s}}}{c} &  =-\frac{c}{\dot{\sigma}}\nabla
\sigma.\label{vs}
\end{align}
In a stationary solution of the form $\Phi\left(  \mathbf{x},t\right)
=e^{-i\omega t}\chi\left(  \mathbf{x}\right)  ,$ we have $\dot{\sigma}
=-\omega.$ Eq. (\ref{vs}) was first derived in Ref.\cite{KXZ}.

\section{Superfluid rotation}

\subsection{Rotating frame}

A superfluid can flow frictionlessly past a wall at low velocities;
dissipations can occur only when the velocity exceeds a critical value
necessary to excite the system. Similarly, a superfluid contained in a
rotating bucket will remain at rest, until the angular frequency of the bucket
exceeds a critical value, whereupon a rotational velocity field occurs through
the creation of quantized vortices, with vortex lines parallel to the axis of
rotation. In a steady state, these vortices form a lattice of specific form.
Theoretical treatment of this problem is best done in a coordinate frame
rotating with the bucket, for it easily exposes the inertial forces, i.e., the
Coriolis force and the centrifugal force, that are responsible for vortex
creation. The actual superfluid we are studying may not be contained in a
bucket, and we may not be in a rotating frame, but the mechanism for vortex
creation may be attributed to a local version of these forces.

Consider a rotating frame with angular velocity $\Omega_{0}$ about the $z$
axis. In spatial spherical coordinates, the lab frame $(t^{\prime},r^{\prime
},\theta^{\prime},\phi^{\prime})$ and the rotating frame $(t,r,\theta,\phi)$
are related by the transformation
\begin{equation} \label{RFT}
t=t^{\prime},\text{ \ \ }r=r^{\prime},\text{ \ \ }\theta=\theta^{\prime
},\text{ \ }\phi=\phi^{\prime}-\Omega_{0}t^{\prime}\text{ .\ }
\end{equation}
The d'Alembertian operator in the rotating frame in flat spacetime is given
by
\begin{equation}
\Box=\Box^{(0)}+R_{\text{Coriolis}}+R_{\text{centrifugal}}
,\label{dalembert}
\end{equation}
where $\Box^{(0)}=\nabla^{2}-\frac{1}{c^{2}}\frac{\partial^{2}}{\partial
t^{2}},$ and
\begin{align}
R_{\text{Coriolis}} &  =\frac{2\Omega_{0}}{c^{2}}\frac{\partial^{2}}{\partial
t\partial\phi},\nonumber\\
~~~R_{\text{centrifugal}} &  =-\frac{\Omega_{0}^{2}}{c^{2}}\frac{\partial^{2}
}{\partial\phi^{2}}.
\end{align}
Eq.(\ref{dalembert}) can be applied to the study of any Klein-Gordon equation in the rotation frame with
interactions, e.g. a self-interaction $V$ which gives the NLKG. Note that the coordinate transformation (\ref{RFT})
is restricted by the condition $\Omega_0 R \ll c$ where $R$ is the radius of system. 
This constraint is satisfied in most rotating systems, from laboratory experiments on liquid $^{4}$He and BEC to neutron stars.  
Here the spacetime background is assumed to be flat. Curved-spacetime cases involving fast-rotating blackholes are studied in \cite{GXCH} and a general treatment will be given in subsection C.

\subsection{Feynman's relation}

Consider $N$ vortices in rotating bucket of radius $R$ and angular velocity
$\Omega.$ At the wall of the bucket, the superfluid velocity is $v_{\text{s}%
}=\Omega R$. Thus, $\nabla\sigma=v_{\text{s}}/\left(  c\xi_{\text{s}}\right)
=\Omega R/\left(  c\xi_{\text{s}}\right) .$ From the relation $
{\displaystyle\oint}
ds\cdot\nabla\sigma=2\pi N$, we obtain
\begin{equation}
\Omega=\pi c\xi_{\text{s}}n_{\text{v}}, \label{feynman}
\end{equation}
where $n_{\text{v}}=N/\left( \pi R^{2}\right)  $ is the number of vortices
per unit area. This formulas can give an estimate of the local vortex density
when the superfluid flows with varying local angular velocity. In
non-relativistic limit $\xi_{\text{s}}\rightarrow\hbar/mc$, we have
$\Omega=\left(  \pi\hbar/m\right)  n_{\text{v}},$ which is called Feynman's relation.
The derivation is valid only for large $N$, since we treat the rotating superfluid 
as if it were a rigid body.

\subsection{Frame-dragging from spacetime geometry}

The transformation to a rotating coordinate frame is equivalent to using the
metric in the following line element:
\begin{align}
ds_{0}^{2} &  =-c^{2}dt^{2}+dr^{2}+r^{2}\sin^{2}\theta d\theta^{2}\nonumber\\
&  +r^{2}(d\phi+\Omega_{0}dt)^{2}.
\end{align}
The cross term $2\Omega_{0}r^{2}d\phi dt$ corresponds to a rotating frame of
angular velocity $\Omega_{0}$. A spacetime-dependent $\Omega_{0}$ may lead to
frame-dragging, i.e., local rotating frames. We give a general treatment of
this effect in the following.

Consider a stationary axially-symmetric background metric with two Killing
vectors $\xi^{a}=\left( \partial/\partial t\right)  ^{a}$ and $\psi
^{a}=\left( \partial/\partial\phi\right)  ^{a},$ which characterize the stationary
condition and the rotational symmetry, respectively. The metric can be
parametrized with coordinates $\{t,u,v,\phi\}$ as follows \cite{Wald}:
\begin{align} \label{wald}
ds^{2} &  =-Adt^{2}+g_{uu}du^{2}+2g_{uv}dudv+g_{vv}dv^{2}\nonumber\\
&  +2Bd\phi dt+Cd\phi^{2},
\end{align}
where the functions $A,B,C$ are related to the Killing vectors:
\begin{align}
A &  =-g_{tt}=-\xi^{a}\xi_{a},\nonumber\\
~~B &  =g_{t\phi}=\xi^{a}\psi_{a},\nonumber\\
~~C &  =g_{\phi\phi}=\psi^{a}\psi_{a}.
\end{align}
In general we should solve the NLKG in the background metric (\ref{wald}) for
rotation problems. Here we consider a small $B$ approximation.
By expanding the d'Alembertian operator in powers of $B,$ we find a cross term
identifiable with the Coriolis force:
\begin{equation}
2g^{t\phi}\frac{\partial^{2}}{\partial t\partial\phi}=\frac{2B}{AC}
\frac{\partial^{2}}{\partial t\partial\phi}+O\left( B^{3}\right)  .
\label{Coriolis}
\end{equation}
Noting that $\Omega=-B/C$ \ is the coordinate angular velocity of locally
non-rotating observers \cite{Wald}, we can write
\begin{equation}
R_{\text{Coriolis}}=\frac{2\Omega}{g_{tt}}\frac{\partial^{2}}{\partial
t\partial\phi}. \label{Cor2}
\end{equation}
In a similar manner we obtain the centrifugal term
\begin{equation}
R_{\text{centrifugal}}=\frac{\Omega^{2}}{g_{tt}}\frac{\partial^{2}}
{\partial\phi^{2}}. \label{centri}
\end{equation}

Examples of metrics with frame-dragging are the BTZ metric in 2+1 dimensional
spacetime, and the Kerr metric in 3+1 dimensional spacetime, which describe
the spacetime curvature around a black hole. Frame-dragging in these metrics
arise from the angular velocity of the black hole. Vortex creation due to this
purely geometric effect is investigated in \cite{GXCH}. From now on we will 
focus on the flat spacetime background and study the NLKG in the rotating frame
defined by the transformation (\ref{RFT}).  

\section{Non-relativistic limit}

\subsection{From NLKG to NLSE}

A solution to the NLKG generally contains frequencies of both signs. A large
frequency $\omega_{0}$ of one sign (say, positive) could develop, due to an
initial field with large positive charge, or a large potential energy, and the
system will approach the non-relativistic limit. We define a wave function
$\Psi$ through

\begin{equation}
\Phi\left(  \mathbf{x},t\right)  =e^{-i\omega_{0}t}\Psi\left(  \mathbf{x}
,t\right)  .\label{Phi_psi}
\end{equation}
Introducing a mass scale $m$ by putting $\omega_{0}=mc^{2}/\hbar$, we find the
equation for $\Psi$:
\begin{equation}
i\hbar\frac{\partial\Psi}{\partial t}=-\frac{\hbar^{2}}{2m}\nabla^{2}
\Psi+U\Psi+\frac{\hbar^{2}}{2mc^{2}}\frac{\partial^{2}\Psi}{\partial t^{2}},
\end{equation}
where $U=\left(  \hbar^{2}/2m\right)  V^{\prime}-\left(  mc^{2}/2\right)  .$
In the limit $c\rightarrow\infty$, we drop the term $\partial^{2}\Psi/\partial
t^{2}$ and obtain the nonlinear Schr\"{o}dinger equation (NLSE).

Let the phase of the non-relativistic wave function be denoted by
$\beta\left(  \mathbf{x},t\right)  :$
\begin{equation}
\Psi=|\Psi|e^{i\beta}.
\end{equation}
The phase of the relativistic field $\Phi$ is thus given by

\begin{equation}
\sigma=-\omega_{0}t+\beta,
\end{equation}
Putting
\begin{equation}
u^{k}\equiv\partial^{k}\sigma=\partial^{k}\beta,
\end{equation}
we obtain, from (\ref{hydro}),

\begin{align}
&  \partial_{t}\left[  \left(  1-\dot{\beta}/\omega_{0}\right)  F^{2}
u^{k}+\frac{F^{2}}{4\omega_{0}}\partial^{k}\partial_{t}\left(  \ln
F^{2}\right)  \right]  \nonumber\\
&  \text{ \ \ \ \ \ \ \ \ \ \ }+\frac{c^2}{\omega_{0}}\partial_{j}
(F^{2}u^{j}u^{k})+\partial^{k}P=0,\nonumber\\
&  \partial_{t}\left[  \left(  1-\dot{\beta}/\omega_{0}\right)  F^{2}\right]
+\frac{c^{2}}{\omega_{0}}\partial_{j}\left(  F^{2}u^{j}\right)  =0,
\end{align}
where the pressure $P$ is defined by
\begin{equation}
P=\frac{1}{2}\left(  F^{2}V^{\prime}-V\right)  -\frac{c^{2}}{4\omega_{0}
}F^2 \,\nabla^{2}\left(  \ln F^{2}\right)  .
\end{equation}
The terms in braces are due to self interaction, while the last term is the
\textquotedblleft quantum pressure" \cite{RMP}. The non-relativistic superfluid
density and velocity can be obtained from (\ref{vs}) by putting $\omega
_{0}=mc^{2}/\hbar$, and formally let $c\rightarrow\infty:$
\begin{align}
\rho_{\text{s}} &  =m|\Psi|^{2}\left(  1+\hbar\dot{\beta}/mc^{2}-v_{\text{s}
}^{2}/c^{2}\right)  +O\left(  c^{-4}\right)  ,\nonumber\\
\mathbf{v}_{\text{s}} &  =\frac{\hbar}{m}\nabla\beta\,\left(  1+\hbar
\dot{\beta}/mc^{2}\right)  +O\left(  c^{-4}\right)  .
\end{align}
From (\ref{hydro}) we obtain the non-relativistic hydrodynamic equations:
\begin{align}
&  \frac{\partial\rho_{\text{s}}}{\partial t}+\nabla\cdot\left(
\rho_{\text{s}}\mathbf{v}_{\text{s}}\right)  =0,\nonumber\\
&  \left(  \frac{\partial}{\partial t}+\mathbf{v}_{\text{s}}\cdot
\nabla\right)  \mathbf{v}_{\text{s}}=-\frac{\nabla P}{\rho_{\text{s}}}- \mathbf{R} .
\end{align}
where $\mathbf{R} = (\nabla \cdot  \mathbf{v}_{\text{s}} + \mathbf{v}_{\text{s}} \cdot \nabla \ln \rho_{\text{s}}) \mathbf{v}_{\text{s}} $ vanishes for an incompressible, divergenceless fluid.

\subsection{NLSE in rotating frame}

We start with the NLKG in a rotating frame:
\begin{align} \label{NLKGrf}
&  \left(  \nabla^{2}-\frac{1}{c^{2}}\frac{\partial^{2}}{\partial t^{2}
}-V^{\prime}+R_{\text{Coriolis}}+~R_{\text{centrifugal}}\right)
\Phi=0,\nonumber\\
&  V^{\prime}=\left(  \frac{mc}{\hbar}\right)  ^{2} - 2m \lambda \, |\Phi|^{2},\nonumber\\
&  R_{\text{Coriolis}}=\frac{2\Omega}{c^{2}}\frac{\partial^{2}}{\partial
t\partial\phi},\nonumber\\
~ &  R_{\text{centrifugal}}=-\frac{\Omega^{2}}{c^{2}}\frac{\partial^{2}
}{\partial\phi^{2}}
\end{align}
where we use a specific form of the potential. Going to the non-relativistic limit
yields
\begin{equation}
i\hbar\frac{\partial\Psi}{\partial t}=\left(  -\frac{\hbar^{2}}{2m}\nabla
^{2}+i\hbar\Omega\frac{\partial}{\partial\phi}+\lambda \,|\Psi|^{2}\right)  \Psi,
\end{equation}
which is the NLSE usually used to describe a rotating BEC \cite{RMP2}. 
The trapping potential can be included as well. 
The centrifugal term does not appear, because it is $-\left(  \hbar^{2}\Omega
^{2}/c^{2}\right)  \partial^{2}/\partial\phi^{2}$, and thus of order $c^{-2}.$
For comparison some features of the relativistic and non-relativistic limit are listed in Table 1.

\begin{table}
\begin{tabular}{c|c|c}
\hline 
\hline
Equations & NLKG & NLSE \\ 
\hline 
Variables & $\Phi ({\bf x}, t) = F e^{ i \sigma} $ & $\Psi ({\bf x}, t) = |\Psi| e^{i \beta}$ \\
\hline 
Superfluid densities $\rho_s$& $-\,\frac{\hbar\dot{\sigma}}{c^{2}\gamma}\,F^{2}$ & $m|\Psi|^{2}$ \\ 
\hline 
Superfluid velocities ${\bf v}_s$ & $-\frac{c}{\dot{\sigma}}\nabla \sigma$ & $\frac{\hbar}{m}\nabla\beta$\\ 
\hline 
Rotating terms & $\frac{2\Omega}{c^{2}}\frac{\partial^{2} \Phi}{\partial
t\partial\phi}$, $-\frac{\Omega^{2}}{c^{2}}\frac{\partial^{2} \Phi}{\partial\phi^{2}}$ & $ i\hbar\Omega\frac{\partial \Psi}{\partial\phi}$ \\ 
\hline 
Feynman relation & $\Omega=\pi c\xi_{\text{s}}n_{\text{v}}$ & $\Omega=\left(  \pi\hbar/m\right)  n_{\text{v}}$  \\
\hline
\hline
\end{tabular} 
\caption{Comparison of some features of the relativistic and non-relativistic limit based on NLKG and NLSE, respectively.}
\end{table}

\section{Current-current interaction with a source}

The complex scalar field $\Phi$ may be coupled to an external source. In
Minkowski spacetime, the only Lorentz-invariant interaction in the Lagrangian
density is the current-current interaction
\begin{equation}
\mathcal{L}_{\text{int}}=-\eta J^{\mu}j_{\mu},
\end{equation}
where $j_{\mu}$ is the conserved current (\ref{current}) of the complex scalar
field, $\eta$ a coupling constant, and $J^{\mu}$ is the conserved current of
the external source. A direct coupling is possible only if the scalar field
has multi-components representing internal symmetry. We can write, more
explicitly,
\begin{equation}
\mathcal{L}_{\text{int}}=-\eta F^{2}J^{\mu}\partial_{\mu}\sigma,
\end{equation}
showing that it is a derivative coupling of the phase of the field. The NLKG
now reads
\begin{equation}
\left(  \Box-V^{\prime}\right)  \Phi-i\eta J^{\mu}\partial_{\mu}\Phi=0.
\end{equation}

Ref. \cite{KXZ} uses an external current of the form
\begin{equation} \label{external}
J^{\mu}  =\left(  \rho,\mathbf{J}\right), ~~~\mathbf{J}   =\rho\mathbf{\Omega\times r,}
\end{equation}
to simulate the presence of a galaxy in a cosmic superfluid. Here,
$\rho(\mathbf{x})$ is the density of the galaxy, represented by a Gaussian
distribution, and $\mathbf{\Omega}$ is its angular velocity. In this case, the
NLKG takes the form
\begin{equation} \label{NLKGcc}
\left(  \Box-V^{\prime}\right)  \Phi+i\eta\rho\left(  \frac{\partial\Phi
}{\partial t}+\mathbf{\Omega}\times\mathbf{r}\cdot\nabla\Phi\right)  =0.
\end{equation}
The last term reads, in a cylindrical coordinate system with azimuthal angle
$\phi$ about the rotation axis,
\begin{equation}
i\eta\rho\mathbf{\Omega}\times\mathbf{r}\cdot\nabla\Phi=i\eta\rho\Omega
\frac{\partial\Phi}{\partial\phi}\,.
\end{equation}
Since $\rho\Omega$ effectively gives a spatially dependent angular velocity,
this term gives rise to an effect similar to frame-dragging. We can compare it
with a Coriolis term
\begin{equation}
\Omega\,\frac{\partial^{2}\Phi}{\partial t\partial\phi}=-i\omega\Omega
\frac{\partial\Phi}{\partial\phi},
\end{equation}
where we have put $\partial\Phi/\partial t=-i\omega\Phi$ for a stationary solution.

\begin{figure}
[htb!]
\begin{center}
\includegraphics[
width=0.50\textwidth, natwidth=1500,natheight=700
]
{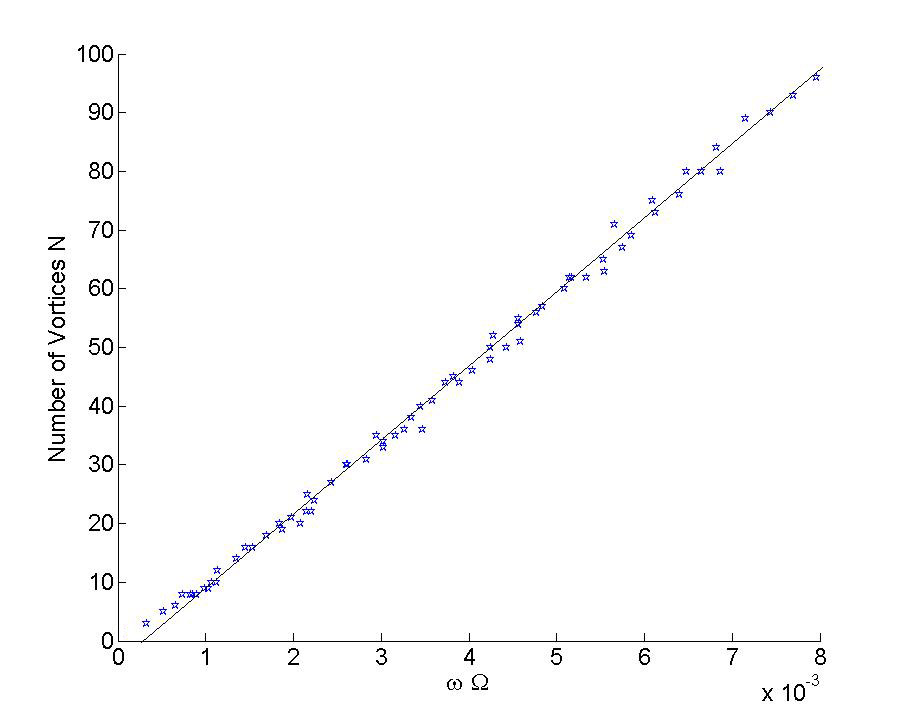}
\caption{The number $N$ of vortices in a rotating bucket as function of the
angular velocity $\Omega$ and the field frequency $\omega$. 
The straight line represents the relativistic Feynman relation (\ref{feynman}), 
which can be rewritten $N=(R^{2}/c^2)\omega\Omega$.
\ Here, $R$ is the radius of the bucket, and $\omega$ is the frequency of the
the stationary solution to the NLKG. }
\end{center}
\end{figure}

\begin{figure}[ptb]
\begin{center}
    \includegraphics[
width=0.40\textwidth, natwidth=1200,natheight=1000
]{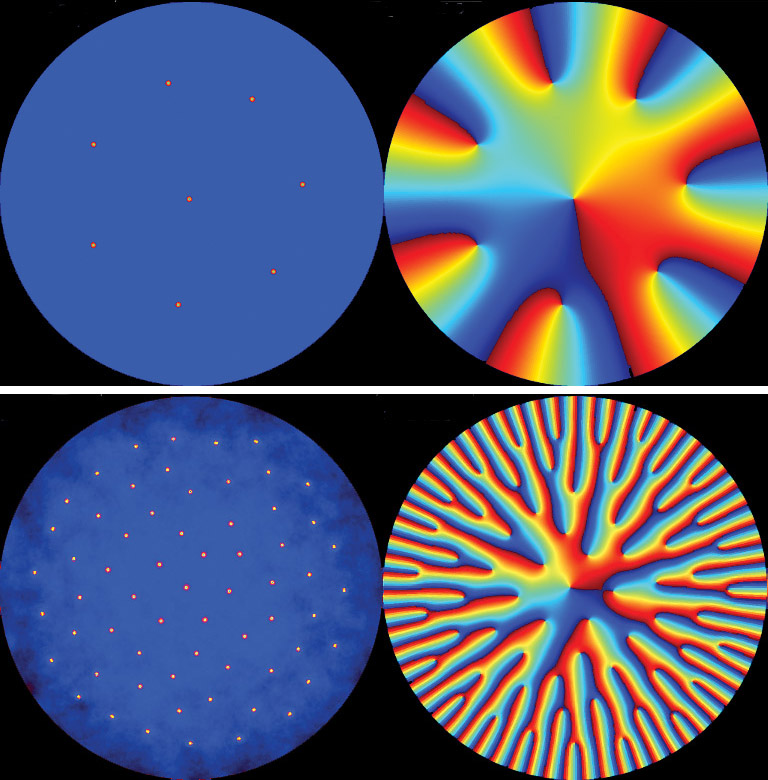}
\caption{Two lattice states with 8 and 63 vortices respectively: The modulus of the field is plotted on the left side and its phase is on the right side. The color map from blue to red (see web version) indicates a phase change of $2\pi$.}
\end{center}
\label{test}
\end{figure}

\section{Numerical Computations}

We solve the NLKG with various inertial or interaction terms in (2+1)- 
and (3+1)-dimensional Minkowski spacetime. Our emphasis here is on vortex
states and vortex dynamics. For the effects of gravity and dark matter
modeling, see Ref.\cite{KXZ}.

The equations are solved by finite difference scheme and spectrum methods are used to 
decouple the discretization for derivatives $\partial_t$ and $\partial_{\phi}$ in the Coriolis term. 
Semi-implicit scheme similar to the Crank-Nicolson method is used for the linear terms. 
The non-linear term is handled explicitly. We impose second order Neumann boundary condition in solving
the NLKG in the rotating frame (\ref{NLKGrf}) and periodic boundary condition in studying the vortex rings lattice and vortex line reconnection with (\ref{NLKGcc}), respectively.

Fig.2 shows in (2+1) dimensions the number of vortices in a rotating bucket as a function of the
angular velocity and the field frequency. The computational data agrees quite well with the
relativistic Feynman relation (\ref{feynman}). As examples Fig.3 shows the modulus (left) and the phase (right) plots of the complex scalar field in stationary states with 8 and 63 vortices, respectively. The locations of vortices are indicated by the dots in the modulus plots, corresponding to positions in the phase plots around which the color changes from blue to red, indicating a phase change of $2\pi$. The vortex distribution is mainly determined by the Coriolis term while some nonuniformity is caused by the centrifugal term. 

In (3+1) dimensions, we introduce an external current of the form (\ref{external}) to
simulate a rotating ``star" immersed in a cosmic superfluid and then solve (\ref{NLKGcc}). 
A vortex-ring solution is found and the magnitude of the modulus gradient is plotted in Fig.4 with the usual rainbow colormap. It shows that the star drags the superfluid 
into local rotation, and there appear vortex rings surrounding the star. As pointed out in Ref. \cite{HLT2, KXZ}, 
the so-called ``non-thermal filaments" observed near the center of the Milky Way may be parts of such
vortex rings around rotating black holes.

\begin{figure*}
[ht!]
\begin{center}
\includegraphics[
natheight=2.00in,
natwidth=6.00in,
height=2.0in,
width=6.0in
]
{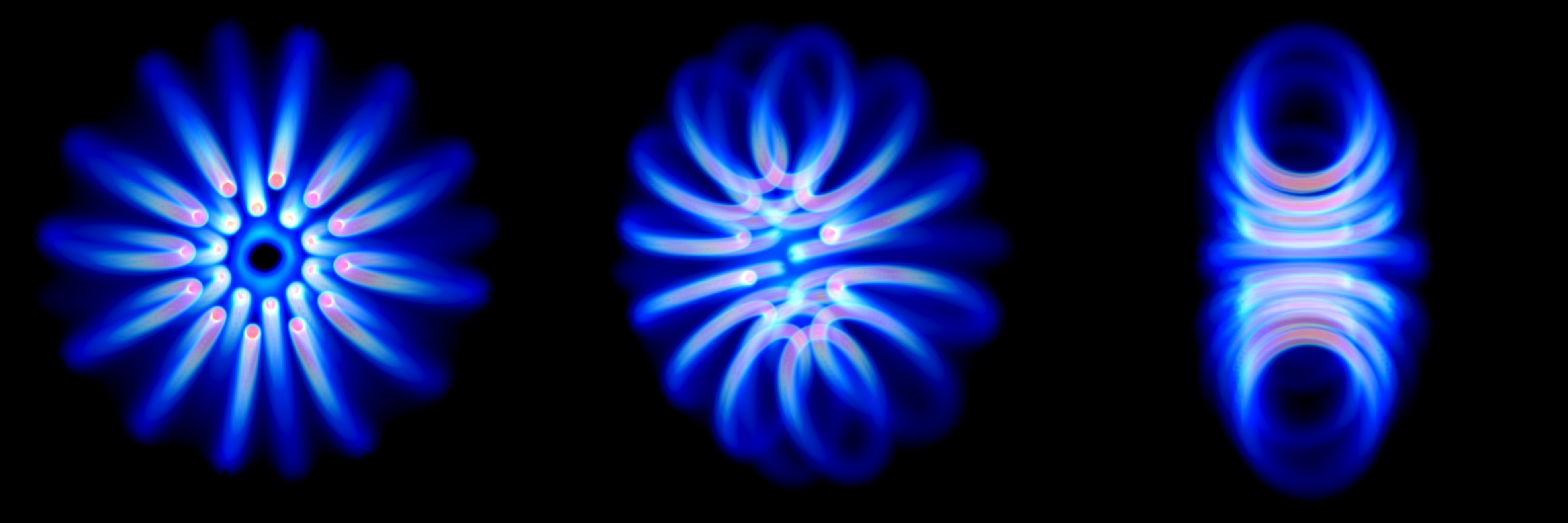}
\caption{System of  3D vortex rings surrounding a rotating ``star", viewed from different perspectives.}
\end{center}
\end{figure*}

Vortex lines in 3D can cross and reconnect. The originally smooth vortex lines
become lines with cusps at the reconnection point, and these cusps spring away
from each other rapidly, creating two jets of energy in the superfluid.
Through repeated reconnections, large vortex rings can be degraded until they
become a vortex tangle of fractal dimensions (quantum turbulence). This
mechanism has been used in \cite{HLT1, HLT2} to create matter in the cosmos during the big
bang era. We can simulate such a reconnection via the NLKG, as shown in Fig.5 by plotting the curl of the phase gradient.
Vortex reconnection in the NLSE has been studied in Ref. \cite{Koplik}. Note that vortex dynamics has also been studied via the local induction approximation (see e.g. \cite{Schwarz:1985zz}) and its connection to the NLSE has been found in \cite{Hasimoto72}. For recent development along this direction, see \cite{Shivamoggi, VanGorder}. 
For actual photographs of vortex reconnection in superfluid helium, see Ref. \cite{Lathrop}.

\begin{figure}
[ptb]
\begin{center}
\includegraphics[
width=0.48\textwidth, natwidth=1500,natheight=680
]
{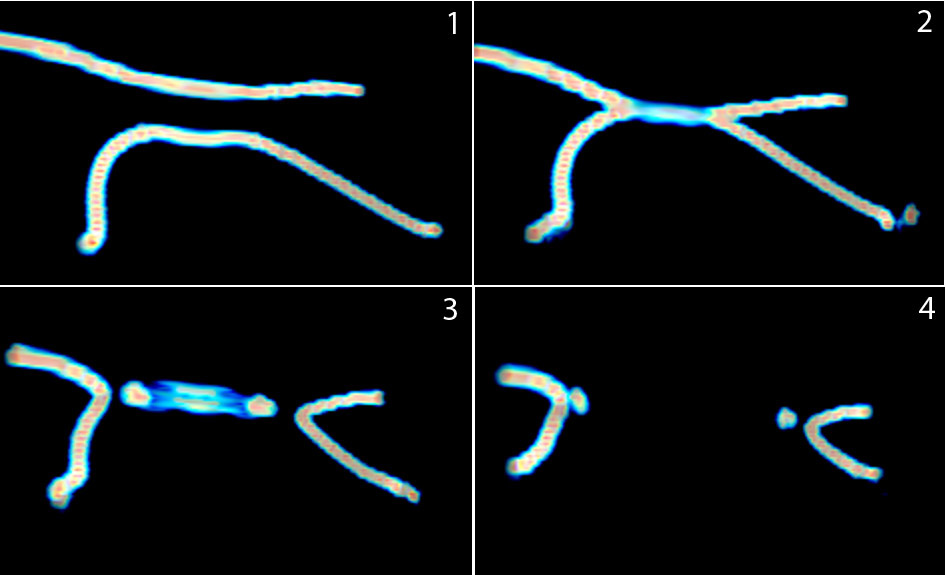}
\caption{Frames 1-4 show the progress of a vortex reconnection, at equal time
steps. An initial configuration of a 3D vortex-line array is constructed with the current-current source term, which  is then turned off to allow free evolution and reconnection of the vortex lines.
}
\end{center}
\end{figure}

\section{Conclusion}

We formulate a framework for investigating superfluidity and the mechanism for creating quantized
vorticity in the relativistic regime, based on a nonlinear Klein-Gordon equation for a complex scalar field. 
This framework is constructed in curved spacetime, then is reduced to flat spacetime and also the non-relativistic limit. 
Our numerical study focuses on the flat-spacetime cases. It shows that quantized vorticity can be created by 
local inertial forces (Coriolis and centrifugal) and current-current interaction. Feynman's relation relating the number of vortices to the angular frequency is numerically verified in the relativistic regime. Our numerical solutions to the NLKG exhibit such features as vortex lattices, 3D vortex rings around rotating stars, and vortex reconnections. In this paper we only consider pure superfluid at zero temperature. It would be interesting to generalize this relativistic formulation to the finite temperature cases in which a normal fluid component and the corresponding effects, such as mutual friction will emerge.

\end{document}